\documentclass[aps,prb,superscriptaddress,floatfix,showpacs,twocolumn]{revtex4}
\usepackage[pdftex]{graphicx}  % needed for figures
\usepackage[update]{epstopdf}
\usepackage{amssymb}
\usepackage{amsmath}
\usepackage{color}
\usepackage{textcomp}
\usepackage[utf8]{inputenc}
\usepackage[normalem]{ulem}
\graphicspath {{./figures/}}
\makeatletter
\def\input@path{{./figures/}}
\makeatother
\usepackage{graphicx}
\usepackage[colorlinks = true,
linkcolor = blue,
urlcolor  = blue,
citecolor = red,
anchorcolor = blue]{hyperref}
\usepackage{amsfonts}
\usepackage{bm}
\usepackage{multirow}
\usepackage{color}
\definecolor{dgreen}{rgb}{0,0.7,0}

\usepackage{tikz}
\usetikzlibrary{shapes,shapes.multipart}

\begin{document}
 \title{Harmonically confined particles with long-range repulsive interactions}
\author{S. Agarwal}
\address{International Centre for Theoretical Sciences, Tata Institute of Fundamental Research, Bengaluru -- 560089, India}
\address{Birla Institute of Technology and Science, Pilani - 333031, India}
\author{ A. Dhar}
\address{International Centre for Theoretical Sciences, Tata Institute of Fundamental Research, Bengaluru -- 560089, India}

\author{M. Kulkarni}
\address{International Centre for Theoretical Sciences, Tata Institute of Fundamental Research, Bengaluru -- 560089, India}

\author{A. Kundu}
\address{International Centre for Theoretical Sciences, Tata Institute of Fundamental Research, Bengaluru -- 560089, India}

\author{S. N. Majumdar}
\address{LPTMS, CNRS, Univ.  Paris-Sud,  Universite Paris-Saclay,  91405 Orsay,  France}

\author{D. Mukamel}
\address{Department of Physics of Complex Systems, Weizmann Institute of Science, Rehovot 7610001, Israel}

\author{G. Schehr}
\address{LPTMS, CNRS, Univ.  Paris-Sud,  Universite Paris-Saclay,  91405 Orsay,  France}

\date{\today}

\begin{abstract}

We study an interacting system of $N$ classical particles on a line at thermal equilibrium. The particles are 
confined by a harmonic trap and repelling each other via pairwise interaction potential that behaves as a power law 
{$\propto \sum_{\substack{i\neq j}}^N|x_i-x_j|^{-k}$} (with $k>-2$) of their mutual distance. This is a generalization of the well known cases of the one component plasma ($k=-1$), Dyson's log-gas ($k\to 0^+$), and the 
Calogero-Moser model ($k=2$). Due to the competition between harmonic confinement and pairwise repulsion, the particles 
spread over a finite region of space for all $k>-2$. We compute exactly the average density profile for large $N$ 
for all $k>-2$ and show that while it is independent of temperature {for sufficiently low temperature}, it has a rich and nontrivial dependence on $k$
with distinct behavior for $-2<k<1$, $k>1$ and $k=1$. 
\end{abstract}

%\pacs{75.50.Lk, 64.60.F-, 02.60.Pn}

\maketitle

\textit{Introduction:} A gas of $N$ classical particles, confined by a harmonic potential on a line and interacting
with each other via pairwise repulsion, constitutes one of the simplest interacting particle systems that have been well studied
in the past. It has seen a recent revival in the wake of the physics of cold atoms.
When the pairwise repulsive interaction decays as a power-law of the distance between the particles, the energy of the
so called Riesz gas~\cite{Riesz}
is {given by}
\begin{equation}
E(\{x_i\})= \frac{1}{2} \sum_{i=1}^Nx_i^2 + \frac{J~\text{sgn}(k)}{2} \sum_{i\neq j} \frac{1}{|x_i-x_j|^k}
\label{energy_k}
\end{equation}
where $J>0$ and $\{x_i\}$ ($i=1,2,\ldots, N$) denote the positions of the particles on the line.
The index $k>-2$ characterizes the strength of the pairwise interaction and 
${\rm{sgn}}(k)$ in the prefactor ensures 
a repulsive interaction. For $k<-2$, the quadratic potential is not strong enough to counter the
strong repulsion and confine the particles. Consequently the particles fly off to $\pm \infty$ and thus the case $k<-2$ is not physically interesting. Given the energy in~\eqref{energy_k}, the joint probability distribution function (PDF) of the particles'
positions is given by the
Boltzmann weight, $P(x_1,\cdots, x_N) = \frac{1}{Z_N(\beta)} e^{-\beta E[\{ x_i\}]}$, where $\beta$ is the inverse
temperature ($\rm k_B = 1$) and $Z_N(\beta)=\int \prod_{i=1}^N {dx_i}\, e^{-\beta E[\{x_i\}]}$ is the normalizing partition function.
The harmonic potential tries to confine the particles
near the center of the trap, while the repulsive interaction tries to push them apart. As a result of the competition 
between the
two terms, it turns out that the particles get confined to a finite region of space for large $N$,
 with a space-dependent average macroscopic density,
$\langle \rho_N(x)\rangle = N^{-1}\, \sum_{i=1}^N \langle \delta(x-x_i)\rangle$ (normalized to unity), where $\langle \ldots \rangle$ 
denotes an average with respect to the Boltzmann weight.
A basic natural question is: what is {the configuration of $x_i$'s that minimizes the energy} in \eqref{energy_k} for large $N$ and what is the
density profile in the ground state~? 
This is a classic and important optimization problem {both in physics (see below) and in mathematics 
(see e.g. Refs.~\onlinecite{Land72,CGZ14,LS17})} whose solution, for generic $k>-2$, 
is hitherto unknown. 
A related question is: how does the average density profile depend on the inverse temperature $\beta$ ?

This problem is of great general interest as there are varied physical systems that correspond to special values of $k$.
We start with $k=-1$ where the interaction is linearly repulsive with distance. This is the well known
one dimensional one-component plasma (1$d$OCP)~\cite{Forrester_book}, consisting of oppositely charged 
particles with pairwise Coulomb interaction (linear in $1d$) and
overall charge neutrality~\cite{Lenard_61,Prager_62,Baxter_63,dhar2017,dhar2018}.
Integrating out the positions of the negative charges gives rise to an effective quadratic confinement for
the positive charges and the effective energy of the $N$ positive charges with 
coordinates $\{x_i\}$ is precisely given by \eqref{energy_k} with $k=-1$. In this case, the energy can
be easily minimized by ordering the positions of the particles leading to an 
equispaced configuration~\cite{Lenard_61,Prager_62,Baxter_63,dhar2017,dhar2018}. 
Moreover, for large $N$, the average density profile $\langle \rho_N(x)\rangle $ turns out to be independent of $\beta$
and approaches a scaling form
$\langle \rho_N(x)\rangle \to (1/N) \tilde \rho_{\rm OCP}(x/N)$, where the scaled density $\tilde \rho_{\rm OCP}(y) = 1/(2J)$
is uniform over the interval $[-J,+J]$ and vanishes outside~\cite{CGZ14,Lenard_61,Prager_62,Baxter_63,dhar2017,dhar2018,Cunden_2017}.

The second and perhaps the most well studied example corresponds to the limit $k\to 0^+$,
where we replace ${\rm sgn}(k)$ in Eq.~(\ref{energy_k}) by $+1$, use $|x_i-x_j|^{(-k)} \approx 1 - k 
\log |x_i-x_j|$ and set $J=1/k$. The energy in \eqref{energy_k} then reduces, up to an overall additive constant, to
\begin{eqnarray}\label{log_gas}
E[\{ x_i\}] = \frac{1}{2} \sum_{i=1}^N x_i^2 - \frac{1}{2} \sum_{i \neq j} \ln |x_i-x_j| \;.
\end{eqnarray}
This is the celebrated log-gas of Dyson~\cite{Dyson_1962}. For the special values of $\beta=1$, $2$ and $4$,
the Boltzmann weight of the log-gas $P(\{x_i\})= \frac{1}{Z_N(\beta)}\, \exp[-\beta E[\{ x_i\}]]$ can be identified with
the joint distribution of $N$ real eigenvalues of an $N\times N$ matrix belonging to
the Gaussian ensembles of the random matrix theory (RMT): respectively Gaussian Orthogonal Ensemble (GOE), Gaussian Unitary  
Ensemble (GUE) and Gaussian Symplectic Ensemble (GSE)~\cite{Mehta_book,Forrester_book}. Gaussian ensembles are
the cornerstones of RMT with myriads
of applications, ranging from nuclear physics, mesoscopic transport, quantum chaos, 
number theory all the way to finance and big-data
science~\cite{Mehta_book,Forrester_book,Akemann_book,LMV_book}.
The log-gas with arbitrary $\beta>0$ also appears in RMT as the joint PDF of the eigenvalues of the so called Dumitriu-Edelman
$\beta$-ensemble of tridiagonal random matrices~\cite{DE_2002}.  
The average density for large $N$
converges to the scaling form, independently of $\beta$, 
\begin{eqnarray} \label{scaling_density}
\langle \rho_N(x) \rangle \approx \frac{1}{\sqrt{N}} \tilde 
\rho_{\rm sc}\left( \frac{x}{\sqrt{N}}\right) \;, \; \tilde \rho_{\rm sc} (y) = \frac{1}{\pi} \sqrt{2 - y^2} \;.
\end{eqnarray}
Thus the scaled density is supported over $[-\sqrt{2},+\sqrt{2}]$ and is known as the celebrated Wigner 
semi-circular law~\cite{Wigner_1951}. This central
result of RMT has been instrumental in understanding the global properties in a variety 
of systems including growth models in $1+1$ dimensions belonging to the KPZ universality 
class~\cite{PS_2000}, non-interacting trapped fermions~\cite{fermions_us1, fermions_us2, fermions_us3} 
where the Wigner semi-circle can also be obtained from the so-called local density approximation (LDA) \cite{LDA}, 
non-intersecting Brownian motions~\cite{fisher_vicious,vicious,BBMP_2014}, 
graph theory and communication networks~\cite{Akemann_book}. 

The third physical example corresponds to $k=2$, i.e., with inverse-square repulsion. For $k=2$, \eqref{energy_k} is
the celebrated Calogero-Moser
model~\cite{calogero1969ground,calogero1971solution,calogero1975exactly,moser1976three} which is integrable and is ubiquitous in
diverse fields \cite{polychronakos2006physics}.
Curiously, it turns
out that for any finite $N$, in the minimum energy configuration of both the log-gas ($k\to 0^+$) and the 
Calogero-Moser model ($k=2$),
the particle positions $x_i$'s coincide exactly~\cite{Calogero_1981,AKD_2019} with the $N$ zeros of the Hermite polynomial of 
degree $N$. Consequently,
for large $N$, the scaled average density for $k=2$ also approaches the Wigner semi-circular law in 
Eq.~(\ref{scaling_density}) and as in the log-gas case, the semi-circular law is independent of $\beta$.
This is rather strange: even though the long-range repulsive interaction in the log-gas is much stronger than that
of the inverse-square gas, the average density profile is identical in the two cases.
%This is rather strange: even though the pairwise interaction in the log-gas and 
%the inverse-square gas are completely different, the average density profile is identical in the two cases! 
This raises a very interesting and natural question:
How does the shape of the average scaled density for large $N$ vary as one tunes the parameter $k$ ?
The appearance of the semi-circular law both for $k\to 0^+$ and $k=2$ suggests an intriguing possibility of a
`non-monotonic' dependence of the shape of the average density as one tunes up $k$.
This question on the dependence of the average density on $k$ also has practical implications in a number of physical contexts. 
For example, in a typical cold atom experimental set-up, a 
quadratic confining potential is natural due to the usage of optical laser traps. In addition, it is possible to induce long-range 
power-law repulsive interactions between such atoms.  For instance, charged particles interacting via the $3d$ Coulomb repulsion, 
but confined to a line by highly anisotropic optical trap, would correspond to $k=1$. Similarly, $k=3$ 
describes a dipolar gas confined to $1d$ \cite{lu2011strongly,griesmaier2005bose,ni2010dipolar}. 
For $k\to \infty$, the Riesz gas reduces to a truely short-ranged repulsive gas similar to the harmonically confined
screened Coulomb or Yukawa-gas studied in Refs.~\onlinecite{Cunden_2017,Cunden_2018}.
In addition, other values of $k$ 
have either been realized in experimental setups or could potentially be realized \cite{brown2003rotational}.

\begin{figure}[t]
\includegraphics[width=\linewidth]{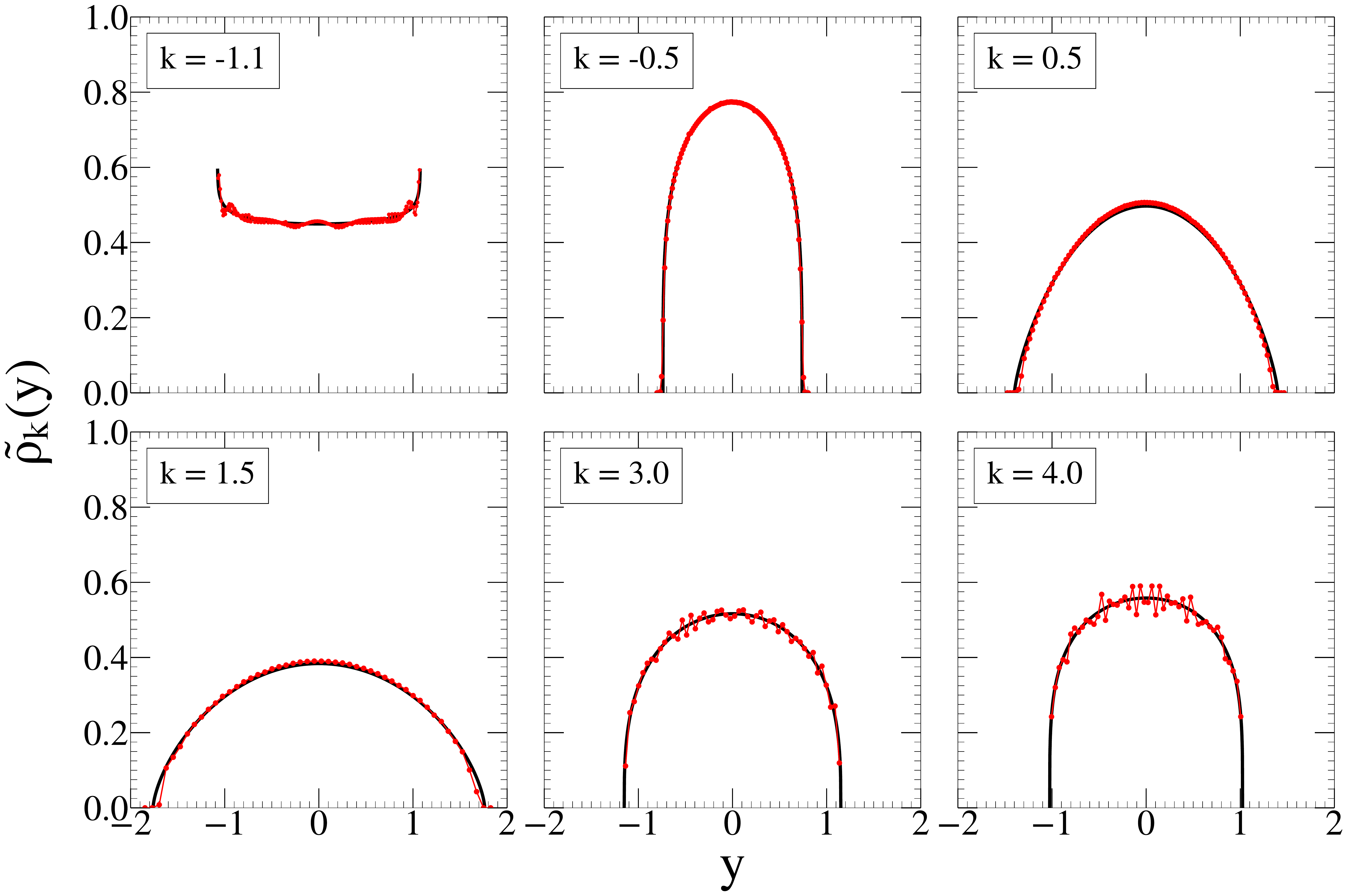}	
\caption {The numerical (MC) average scaled density $\tilde \rho_k(y)$ vs $y$ (red dots) for 
different values of $k$: for $-2<k<1$ in the top
three panels and 
for $k>1$ in the bottom three panels (where $N = 200$ and $\beta = 2$). The numerical curves are compared to
analytical predictions (black) with excellent agreement. 
Oscillations are somewhat prominent at higher $|k|$ due to the finite $N$ effects. The ensemble average is over $2\times 10^8$ MC samples.}% \textcolor{blue}{Can you produce a new figure with a modified $y$-label $\tilde \rho_k(y)$ and larger characters?}} 
\label{fig1} 
\end{figure}

In this Letter, we address this interesting question of the dependence of the average density on $k$ and
obtain exact results for large $N$. We show that for large $N$ the average density is independent of $\beta$ for all
$k$ ({for sufficiently large $\beta$}) \cite{footnote_beta}, but has a rich and nontrivial dependence on $k$.
On general grounds the average density is expected to have 
a scaling form $\langle \rho_N(x)\rangle \approx N^{-\alpha_k} \tilde \rho_k(x/N^{\alpha_k})$ for large $N$, where $N^{\alpha_k}$ corresponds
to the typical scale of the position of the particles in the trap. Indeed, we do find this behavior, but with a 
twist. We show that
there is a drastic change of behavior of the exponent $\alpha_k$ as well as the scaling function $\tilde \rho_k(y)$ at $k=1$. For the 
exponent, we get 
\begin{eqnarray}\label{alpha_k}
\alpha_k = 
\begin{cases}
&\frac{1}{k+2} \;, \;\quad\quad -2<k<1 \\
&\frac{k}{k+2} \;, \quad\quad\quad k > 1 \;.
\end{cases}
\end{eqnarray} 
The scaling function $\tilde \rho_k(y)$ has a support over $[-\ell_k/2, + \ell_k/2]$ and can be expressed as $\tilde \rho_k(y) = \ell_k^{-1} F_k(y/\ell_k)$ where $F_k(z)$ is given by
\begin{eqnarray}\label{F_k}
F_k(z) = \frac{1}{B(\gamma_k + 1, \gamma_k+1)} \left(\frac{1}{4} - z^2 \right)^{\gamma_k}  \;.
\end{eqnarray}
with $-1/2\leq z \leq 1/2$ and $B(a,b)$ is the standard Beta function. Thus the density either diverges or 
vanishes at the 
two scaled edges $z = \pm 1/2$ with an exponent $\gamma_k$ which also exhibits a change of behavior at $k=1$, namely
\begin{eqnarray}\label{gamma_k}
\gamma_k =
\begin{cases}
&\frac{k+1}{2} \;, \;\quad\quad -2<k<1 \\
&\frac{1}{k} \;, \quad\quad\quad k > 1 \;.
\end{cases}
\end{eqnarray}
The support length $\ell_k$ is non-universal and depends explicitly on $k$ and the coupling strength $J$ (for the exact expressions 
of $\ell_k$, see Eqs. (36) and (65) of Supp. Mat.~\cite{suppl}). The scaling function $F_k(z)$ depends only on $k$, and is independent of $\beta$ and $J$.
The case $k=1$ is marginal with additional logarithmic corrections (we discuss this later). We show that this change of 
behavior at $k=1$ can be traced back to the fact that, for $k<1$, the large distance behavior of the interaction term controls the 
large $N$ behavior of the density. In contrast, for $k>1$, the limiting density is determined by the short distance behavior of the 
interaction term. This gives rise to an effective field-theory that is fundamentally different for $k<1$ and $k>1$.  
Thus $k\to 0^+$ (log-gas) and $k=2$ (inverse-square gas) share the same average density profile, but the physics
is rather different in the two cases. For $k\to -1$, we recover the flat density of the 1$d$OCP. Also, in the limit $k\to \infty$
we again get a flat density, consistent with the results for
the $1d$ harmonically confined Yukawa gas~\cite{Cunden_2017,Cunden_2018}.
We also performed Monte-Carlo (MC) simulations for several values of $k$, finding excellent agreement with our analytical 
predictions (see Fig.~\ref{fig1}).
 
\textit{Regime 1: $-2<k<1$:} Assuming both terms in the energy \eqref{energy_k} are of the same 
order for large $N$, the energy scale can be estimated as follows.
%In this regime, both terms in the energy in \eqref{energy_k} are of the same 
%order for large $N$ and can be estimated as follows. 
Let the typical position of a particle {scales} 
as $x_{\rm typ}\sim N^{\alpha_k}$ for large $N$, where $\alpha_k$ is to be determined. 
Then the first term in \eqref{energy_k} scales as $\sim N^{2\alpha_k+1}$, while the second term (where the double sum
contains typically $N(N-1)\approx N^2$ terms) scales as $\sim J\, N^{2-k\, \alpha_k}$. Demanding they are of the same order
fixes the exponent $\alpha_k= \frac{1}{k+2}$ (see the first line in Eq.~(\ref{alpha_k})).
Hence the total energy 
scales as $E \sim N^{2\alpha_k +1} \sim N^{\frac{4+k}{2+k}}$ in this regime. 
To find the configuration that dominates the partition function $Z_N(\beta)$ for 
large $N$, we generalise the method used for the log-gas ($k\to 0^+$ limit)~\cite{Dyson_1962,DM_2006,DM_2008,Saff_book}.
It turns out to be convenient to express the coarse grained energy in terms of a macroscopic density
$\rho_N(x)$ and use the relation 
$\sum_{i=1}^N f(x_i)\approx N\, \int f(x)\, \rho_N(x)\, dx$, valid for any smooth function $f(x)$. Next, we 
rescale $x= N^{\alpha_k}\, y$ with $y\sim O(1)$. Under this rescaling, the density transforms as
$\rho_N(x) \approx N^{-\alpha_k} {\tilde \rho_k}(y= x\, N^{-\alpha_k})$, where we assume
$\tilde \rho_k(y)$ is smooth and normalizable,
$\int{\tilde \rho_k}(y) dy=1$. Consequently, the coarse grained partition function for large $N$ can be 
expressed as a functional integral over the density field $\tilde \rho_k(y)$ (for details see Supp. Mat.~\cite{suppl}) 
\begin{eqnarray}\label{Z_1}
Z_N(\beta) \sim \int d \mu \int \mathcal{D}[\tilde \rho_k] 
\exp \left(-\beta \, N^{\frac{4+k}{2+k}} \Sigma[\tilde \rho_k(y)] \right)
\end{eqnarray}
where the action $\Sigma[\tilde \rho_k(y)]$ is given by (see also Ref.~\onlinecite{Serfaty_book} for a rigorous proof in the cases $0<k<1$)
\begin{eqnarray}\label{Sigma}
\Sigma[\tilde \rho_k(y)]&=&  \frac{J~\text{sgn}(k)}{2} \int dy\int dy'~\frac{\tilde \rho_k(y)\tilde \rho_k(y')}{|y-y'|^k}\\
&+&\frac{1}{2} \int dy~y^2\tilde \rho_k(y)  -\mu \left(\int dy~\tilde \rho_k(y)-1 \right) \, . \nonumber 
\end{eqnarray}
Here $\mu$ is the Lagrange multiplier that enforces the constraint $\int \tilde \rho_k(y) dy=1$. 
Note that in the integrand of Eq.~(\ref{Z_1}), we have only kept the leading order contributions to the
energy. Both the entropy term (generated in going from microscopic configurations to the macroscopic
density) as well as the short distance behavior of the interaction energy have been neglected,
as they are of lower order in $N$ for $-2<k<1$. {This is valid as long as $\beta \gg N^{-2 \alpha_k}$ where $\alpha_k = 1/(k+2)$ \cite{footnote_beta,suppl}.} Thus the
effective action $\Sigma[\tilde\rho_k(y)]$ is manifestly {\em non-local} reflecting the long-range nature of the
repulsive interaction. We will see later that this non-locality manifests only for $-2<k<1$.
For large $N$, the partition function in Eq.~(\ref{Z_1}) can then be evaluated by the saddle point method.
Minimizing the action $\Sigma[\tilde\rho_k(y)]$ in \eqref{Sigma} with respect (w.r.t.) to $\tilde \rho_k(y)$ 
gives the saddle point equation for the optimal density
\begin{equation}
\frac{y^2}{2} + {J\,\text{sgn}(k)}\int dy'~\frac{\tilde \rho_k(y')}{|y'-y|^k} = \mu\, . 
\label{sadl-eq-1}
\end{equation}
This equation is valid over the support of $\tilde \rho_k(y)$. The density is clearly symmetric in
$y$, hence the support is over $[-\ell_k/2, \ell_k/2]$ where $\ell_k$ is fixed using the normalization {$\int_{-\ell_k/2}^{\ell_k/2} \rho_k(y) dy=1$}.
Taking a further derivative of \eqref{sadl-eq-1} w.r.t. $y$ leads to a singular integral equation
\begin{eqnarray}\label{sp_eq}
PV\, \int_{-\ell_k/2}^{\ell_k/2} \frac{\text{sgn}(y'-y)}{|y-y'|^{k+1}} \tilde \rho_k(y')~dy'= -\frac{y}{J\,|k|},\, k\neq 0
\end{eqnarray}
where $PV$ denotes the principal value which needs to be taken only for $k>0$. Also, for $k\to 0$, $J$ has to 
be rescaled such that
$J\, |k|=1$. Solving this singular integral equation poses the main technical challenge.
Fortunately, it turns out that for $-2<k<1$, this equation can be transformed into the well studied Sonin form
~\cite{Sonin,Popov_1982,Widom_1999,Buldyrev_Sonin,Derrida_2007,Cividini_2017,Asaf_2019}, and can subsequently
be inverted to obtain $\tilde \rho_k(y)$ explicitly~\cite{suppl}. We then obtain
the exact saddle point density $\tilde \rho_k(y)=\ell_k^{-1} F_k(y/\ell_k)$ where $\ell_k$ is given
in Supp.~Mat.~\cite{suppl} and $F_k(z)$ is given
in Eq.~(\ref{F_k}) with $\gamma_k=(k+1)/2$. Note that the Sonin inversion formula also indicates that
there is no physical solution (saddle point density) for $k>1$. Hence, this solution is valid only in the range
$-2<k<1$. Furthermore, since $\beta$ appears only in the factor $\beta\, N^{(4+k)/(2+k)}$ outside the action
$\Sigma[\tilde \rho_k(y)]$ in \eqref{Z_1}, it is clear that the saddle point density is independent of $\beta$:
large $N$ is equivalent to large $\beta$. In addition, the average density $\langle \rho_N(x)\rangle$, for large $N$,
clearly coincides with the saddle point density as the average over all possible densities is 
dominated by the saddle point.
In Fig.~\ref{fig1} upper panels, we compare our theoretical predictions with MC simulations for three
representative values of $k$ in the range $-2<k<1$ and find excellent agreement. Note that in the range 
 $-2<k<-1$ the density diverges at the edges $\pm l_k/2$, while for $-1<k<1$ the density vanishes at the edges. Exactly
at $k=-1$, the density is flat, consistent with the 1$d$OCP result. 

\textit{Regime 2: $k\geq1$}:
It turns out 
that, for $k > 1$, the interaction term in \eqref{energy_k} containing the double sum is dominated by 
particles which are very close to each other, i.e., almost nearest neighbours.
As a result, the short distance properties of the interaction term plays a more dominant 
role compared with its long distance behavior. This leads to an effective field theory which is
 {\em local} in the density and is much simpler. To compute the effective coarse grained energy
for large $N$, we then take a different path than the $k<1$ case (for details see Supp. Mat.~\cite{suppl}). 
First, it is convenient to order the particle positions so that $x_i$ increases with the label $i$ (this
is fine since the energy \eqref{energy_k} is invariant under permutation of labels). We then replace the discrete particle
label $i$ by a continuous coordinate and the position $x_i$ is approximated by a smooth continuous function $x(i)$.
Next, we approximate $x_i-x_j\approx (i-j) x'(i)$ where $x'(i)=dx(i)/di$ and we have kept only the first term
in the Taylor expansion anticipating that it captures the leading short distance behavior. Our next step
is to express $x'(i)$ in terms of the local smooth macroscopic density $\rho_N(x)$ (normalized to unity). 
In fact, the local slope of the smooth function $x(i)$, i.e., $x'(i)>0$ is simply the inverse of the number density
$N\rho_N(x)$, i.e., $x'(i)= 1/(N\rho_N(x))$. Thus the double sum in \eqref{energy_k} 
can be approximated, to leading order for large $N$, by $\sum_{i\ne j} |x_i-x_j|^{-k}= \sum_{i\ne j} |i-j|^{-k}
[N \rho_N(x_i)]^{k}$. The sum over $j$, for fixed $i$, is convergent for all $k>1$ and simply gives a factor 
$2\,\zeta(k)$, where 
$\zeta(k)=\sum_{n=1}^{\infty} n^{-k}$ is the Riemann zeta function. Furthermore, the sum over $i$ can be replaced
by an integral using the relation $\sum_{i=1}^N f(x_i) \approx N \int f(x) \rho_N(x) dx$ mentioned before. Using this
relation in both terms of \eqref{energy_k} leads to a coarse grained energy ${\cal E}\equiv{\cal E}[\rho_N(x)]$ \cite{suppl}
\begin{equation}
{\cal E} \approx \frac{N}{2} \int dx\, x^2 \rho_N(x)
+ J\zeta(k) N^{k+1} \int dx\, \left[\rho_N(x)\right]^{k+1} \;, 
\label{energy_coarse}
\end{equation}
which is completely {\em local} in the density $\rho_N(x)$, unlike \eqref{Sigma} for $-2<k<1$ that involved densities
at two space-separated points. 
We then rescale $x\to x\, N^{-\alpha_k}$ and 
write $\rho_N(x)= N^{-\alpha_k} \tilde \rho_k(y= x\, N^{-\alpha_k})$. It is easy to see 
that for both terms in \eqref{energy_coarse} to be of the same order in $N$ for large $N$, we need to choose
$\alpha_k= k/(k+2)$, as stated in the second line of \eqref{alpha_k}. Hence, the total energy scales 
as $E\sim N^{\frac{3k+2}{k+2}}$ for large $N$. The coarse grained partition function $Z_N(\beta)$ can
then be written as
\begin{eqnarray}\label{Z_2}
Z_N(\beta) \sim \int d \mu \int \mathcal{D}[\tilde \rho_k]
\exp \left(-\beta \, N^{\frac{3k+2}{k+2}} \Sigma[\tilde \rho_k(y)] \right)
\end{eqnarray}
where the action $\Sigma[\tilde \rho_k(y)]$ is given by   
\begin{eqnarray}
 \Sigma[\tilde \rho_k(y)]&=&  \frac{1}{2}\int dy\, y^2 \tilde \rho_k(y)\nonumber 
 +J \zeta(k) \int dy\,\left[\tilde \rho_k(y)\right]^{k+1} \\ & - & \mu \left(\int dy~\tilde \rho_k(y)-1 \right) 
\label{action_coarse} 
\end{eqnarray}
with $\mu$ again denoting the Lagrange multiplier enforcing the normalization of the density. 
Note that we have kept the leading order contribution for large $N$ in the integrand in \eqref{Z_2} and
again neglected the entropy as well as subdominant singular terms which are of lower order in~$N$.
For large $N$, the integral \eqref{Z_2} can again be evaluated
by the saddle point method. Minimizing the action gives the saddle point equation
\begin{equation}
\frac{1}{2} y^2 + J\, \zeta(k)\, (k+1) \left[\tilde \rho_k(y)\right]^{k}= \mu \, .
\label{saddle_point.1}
\end{equation}
Trivially solving this equation gives,
$\tilde \rho_k(y)= (1/\ell_k)\, F_k(y/\ell_k)$, with support over $[-\ell_k/2,\ell_k/2]$ where $\ell_k=2\sqrt{2\mu}$ is 
fixed from the normalization and is given explicitly 
in Supp. Mat.~\cite{suppl}.
The scaling function $F_k(z)$ is then of the form in \eqref{F_k} with the exponent $\gamma_k=1/k$. 
The saddle point density coincides with the average density $\langle \rho_N(x)\rangle$
for large $N$. In addition, since $\beta$ appears only in the combination $\beta N^{(3k+2)/(k+2)}$ outside 
the action in \eqref{Z_2}, clearly the saddle point density and hence the average density $\langle \rho_N(x)\rangle$ is 
independent of $\beta$ for large $N$, {as long as $\beta \gg N^{-2 \alpha_k}$ where $\alpha_k = k/(k+2)$ \cite{footnote_beta,suppl}.} This analytical
prediction is then verified in MC simulations 
(see the bottom panels in Fig.~\ref{fig1}).
 
The marginal case $k=1$ lies at the borderline between Regime 1 and Regime $2$. In this case, one would expect
logarithmic corrections. Indeed, we find~\cite{suppl} that the average density approaches a scaling form,
$\langle \rho_N(x) \rangle \sim L_N^{-1} \tilde \rho_1(x/L_N)$, where the typical position of a particle
scales as $L_N\approx (N\,\ln N)^{1/3}$ for large $N$. The scaling function $\tilde \rho_1(y)$ is supported over 
$[-\ell_1/2,\ell_1/2]$ with $\ell_1=2\sqrt{2\mu}$ and is given~by
\begin{equation}
\tilde \rho_1(y) =  \frac{1}{4J} (2 \mu - y^2),\quad \mu =  \frac{1}{2}\, (3J)^{2/3} \, .
\label{k1case}
\end{equation}
This can be also cast in the scaling form $\tilde \rho_1(y)= (1/\ell_1)\, F_1(y/\ell_1)$ where $F_1(z)$ is given
in \eqref{F_k} with $\gamma_1=1$. Numerical simulations are in good agreement with our analytical prediction,
as shown in the Supp. Mat.~\cite{suppl}. 

\textit{Conclusions:} In this Letter, we have computed analytically the average density profile of a classical gas of
harmonically confined particles that repel each other with the repulsive interaction
behaving as a power-law with exponent $-k$ of the distance between any pair of particles.
Our result generalizes in a nontrivial way, to arbitrary $k>-2$, the three famous classical
examples: the $1d$OCP ($k=-1$), the Dyson's log-gas in RMT ($k\to 0^+$) and the Calogero-Moser model ($k=2$).
We have shown that the underlying effective field theory that determines the average density profile for large $N$
is governed by fundamentally different physics for $-2<k<1$ and $k>1$. In the former case, the large distance
behavior of the interaction potential dominates, while the latter case is governed by its short distance
behavior. 
It would be interesting to study other observables beyond the average density for general $k>-2$.
For instance, for the log-gas ($k\to 0^+$) the position of the rightmost particle $x_{\rm max}$ (the largest eigenvalue
of a random matrix), centered and scaled, is known to converge to the celebrated Tracy-Widom 
distribution~\cite{Tracy_Widom}.
The corresponding extreme value distribution for $k=-1$ has recently been computed exactly~\cite{dhar2017,dhar2018} and the case $k=2$ has been recently computed numerically~\cite{AKD_2019}. It would be interesting to compute the limiting distribution of $x_{\rm max}$ for generic $k>-2$. Finally, it would be interesting to see if our predictions for the average density can be measured in cold atom experiments. From that perspective, it would be nice to
extend our results for the density profile to higher dimensions. 

%Note: After completion of this work, we came to know of a recent Ref.\cite{Riesz_Hardin2018}  in mathematics where the case $ k \geq 1 $ was considered--we thank Ofer Zeitouny for pointing it to us.

\textit{Note added in proof}: After submission of the work, we came to know from O. Zeitouni that the case
$k>1$ was also studied recently in the mathematics literature \cite{HLSS18}. 

\textit{Acknowledgments:} We would like to thank T. Lebl\'e, E.~Saff and S. Serfaty for pointing out  useful references and O. Zeitouni for stimulating discussions. MK would like to acknowledge support from the project 6004-1 of the 
Indo-French Centre for the Promotion of Advanced Research (IFCPAR),  
the Ramanujan Fellowship SB/S2/RJN-114/2016 and the SERB Early Career Research Award ECR/2018/002085 from 
the Science and Engineering Research Board (SERB), Department of Science and Technology, Government of India.
AD, AK, SNM and GS would like to acknowledge support from the project 5604-2 of the Indo-French Centre for the 
Promotion of Advanced Research (IFCPAR). This work was supported by a research grant from the 
Center for Scientific Excellence at the Weizmann Institute of Science. 
AD, AK, SNM and GS acknowledge the hospitality of the Weizmann Institute during the SRITP 
Workshop ``Correlations, Fluctuations and Anomalous Transport in Systems Far from 
Thermal Equilibrium'' held at the Weizmann Institute in January 2018. 
SNM acknowledges the hospitality of the Weizmann Institute
during a visit as a Weston Professor in 2019 and the support from the Science
and Engineering Research Board (SERB, government
of India), under the VAJRA faculty scheme (Ref.
VJR/2017/000110) during a visit to Raman Research Institute,
where part of this work was carried out.  
We would like to thank the ICTS program ``Universality in random structures: 
Interfaces, Matrices, Sandpiles  (Code: ICTS/URS2019/01)" for enabling valuable discussions with many participants. AK acknowledges support from DST 
grant under project No. ECR/2017/000634.

%\bibliography{references.bib}

\end{document}

\title{\underline{Supplementary Material} to ``Harmonically confined particles with long-range 
repulsive interactions"}
\author{S. Agarwal}
\address{International Centre for Theoretical Sciences, Tata Institute of Fundamental Research, Bengaluru -- 560089, India}
\address{Birla Institute of Technology and Science, Pilani - 333031, India}
\author{ A. Dhar}
\address{International Centre for Theoretical Sciences, Tata Institute of Fundamental Research, Bengaluru -- 560089, India}

\author{M. Kulkarni}
\address{International Centre for Theoretical Sciences, Tata Institute of Fundamental Research, Bengaluru -- 560089, India}

\author{A. Kundu}
\address{International Centre for Theoretical Sciences, Tata Institute of Fundamental Research, Bengaluru -- 560089, India}

\author{S. N. Majumdar}
\address{LPTMS, CNRS, Univ.  Paris-Sud,  Universite Paris-Saclay,  91405 Orsay,  France}

\author{D. Mukamel}
\address{Department of Physics of Complex Systems, Weizmann Institute of Science, Rehovot 7610001, Israel}

\author{G. Schehr}
\address{LPTMS, CNRS, Univ.  Paris-Sud,  Universite Paris-Saclay,  91405 Orsay,  France}

\date{\today}

%\pacs{75.50.Lk, 64.60.F-, 02.60.Pn}
\maketitle

\section{Continuum approximation for the interaction term}

In Eq. (1) of the main text, we defined the energy of a microscopic configuration $\{x_i\}$ as
\begin{equation}
E[\{x_i\}]= \frac{1}{2} \sum_{i=1}^N x_i^2 + \frac{J\, {\rm sgn}(k)}{2} \sum_{i\ne j} \frac{1}{|x_i-x_j|^k}\, ,
\label{energy_SM2}
\end{equation}
where we assume $k>-2$. The partition function is defined as $Z_N(\beta)= \int \prod_{i=1}^N dx_i\, 
e^{-\beta\, E[\{x_i\}]}$. It turns out to be convenient to order the particle positions $x_i$'s such that
$x_i$ increases monotonically with the label $i$. Since the energy $E[\{x_i\}]$ in \eqref{energy_SM2} is invariant
under any permuation of the labels, it follows simply that
\begin{equation}
Z_N(\beta)= N!\, \underset{x_1<x_2<\ldots< x_N}{\int\int\dots\int} dx_1\, dx_2\dots\, dx_N\, e^{-\beta\, E[\{x_i\}]}\, .
\label{pfo_SM2}
\end{equation} 
Evaluating this microscopic $N$-fold integral in \eqref{pfo_SM2} exactly for arbitrary $N$ is hard. 
However, one can make progress in the large $N$ limit where one can 
coarse grain the system in terms of a smooth macroscopic density profile $\rho_N(x)$.
Then the idea is to perform the integral in \eqref{pfo_SM2} in two steps. First, 
one fixes the macroscopic non-negative density $\rho_N(x)$ (normalized to unity),
and integrates over
all microscopic configurations in the partition sum \eqref{pfo_SM2} compatible with this macroscopic density $\rho_N(x)$.
Secondly, one integrates over all possible smooth macroscopic density profiles 
$\rho_N(x)$--leading to a functional integral over $\rho_N(x)$. 

It turns out (as we show below) that the
energy in \eqref{energy_SM2} can be expressed as a functional over only this macroscopic density, i.e.,
$E[\{x_i\}] \approx {\cal E}[\rho_N(x)]$, to leading order for large $N$. Then, the partition function in \eqref{pfo_SM2}
can be approximated~as
\begin{equation}
Z_N(\beta) \approx N! \, \int {\cal D}[\rho_N(x)]\, {\cal N}[\rho_N(x)]\, e^{-\beta\, {\cal E}[\rho_N(x)]}\, 
\delta\left( \int dx\, \rho_N(x)-1\right)\, 
\label{pfo_cg_SM2}
\end{equation}
where the delta function enforces the normalization of the macroscopic density (to unity) and ${\cal N}[\rho_N(x)]$ 
is a combinatorial factor that
counts the number of microscopic configurations compatible with a given macroscopic profile $\rho_N(x)$.
Hence ${\cal N}[\rho_N(x)]= \exp\left[ S[\rho_N(x)]\right]$, where $S[\rho_N(x)]$ is the entropy associated 
with the macroscopic 
configuration $\rho_N(x)$. One can explicitly compute the entropy term and show that
$ S[\rho_N(x)] = -N\, \int dx\, \rho_N(x)\, \ln [\rho_N(x)] $  up to an additive constant (for a simple 
derivation in the context of the log-gas, see e.g. Refs. \onlinecite{DM_2008,ABMV_2013}).
Furthermore, the delta function enforcing the normalization constraint can be replaced by its integral
representation, $\delta(y)= \int \frac{d\mu}{2\pi}\, e^{-\mu y}$  where the integral is along the imaginary $\mu$ axis.
Putting all these together on the right hand side (rhs), we obtain, up to an overall (unimportant) multiplicative factor,
\begin{equation}
Z_N(\beta) \sim  \int d\mu\, \int {\cal D}[\rho_N(x)]\, \exp\left[- \beta {\cal E}[\rho_N(x)]-
\mu\left(\int \rho_N(x)\, dx-1\right) - N\, \int dx\, \rho_N(x)\, \ln [\rho_N(x)]\right]\, .
\label{pfo_cg1_SM2}
\end{equation}
It will turn out that for large $N$ and fixed $k>-2$, the energy term will scale as 
$N^{b_k}$ with $b_k>1$, compared to the entropy term ($\sim O(N)$). Hence the energy dominates 
for all $k>-2$ and one can ignore the entropy term for large $N$. Ignoring the entropy term leads to
\begin{equation}
Z_N(\beta) \sim  \int d\mu\, \int {\cal D}[\rho_N(x)]\, \exp\left[- \beta {\cal E}[\rho_N(x)]
-\mu\left(\int \rho_N(x)\, dx-1\right)\right]\, .
\label{pfo_cg2_SM2}
\end{equation}

Thus, our next task is to show that indeed the energy $E[\{x_i\}]$ can be approximated by 
${\cal E}[\rho_N(x)]$, a functional
of only the macroscopic density $\rho_N(x)$. This is far from obvious, in particular for the interaction
term (the second term) on the rhs of \eqref{energy_SM2}. We now show how it can be done. It turns
out that the form of ${\cal E}[\rho_N(x)]$ is very different for the three cases: (i) $-2<k<1$ (ii) $k>1$ and (iii) $k=1$.
To proceed, the first step is to approximate the
microscopic ordered coordinate $x_i$ by a smooth monotonically increasing function $x(s)$ in the large 
$N$ limit, such that $x(i)=x_i$. 
One can relate this smooth function $x(s)$ to the local macroscopic density by noting that
\begin{equation} 
N\, \int^{x}\rho_N(x')dx'= s(x)\, ,
\label{density_relation}
\end{equation}
which follows from the fact that the total number of particles up to label $s$ is simply $s$.
Taking a further derivative of \eqref{density_relation} with respect to $x$ yields
\begin{equation}
\frac{dx}{ds} =  \frac{1}{N \rho_N (x(s))} \, .
\label{slope_density_SM2}
\end{equation}
Note that this relation \eqref{slope_density_SM2} is true only for ordered configurations such that $dx/ds>0$.
This is needed since the rhs of \eqref{slope_density_SM2} is positive because the density is manifestly positive.
The average of any function $f(x_i)$ of 
the coordinates $x_i$, can then be expressed in terms of the macroscopic density as 
\begin{equation}
\sum_{i=1}^N f(x_i) \approx \int f(x) ds = N\, \int f(x)\, \rho_N(x)\, dx \, ,
\label{identity_SM2}
\end{equation}
which holds for any smooth function $f(x)$. For example,
using $f(x)=x^2$ in \eqref{identity_SM2}, the first term (corresponding to the external harmonic
potential) on the rhs of \eqref{energy_SM2} can be simply expressed as a functional of $\rho_N(x)$
\begin{equation}
{\cal E}_{\rm harmonic}[\rho_N(x)]\approx \frac{N}{2} \int x^2\, \rho_N(x)\, dx \, ,
\label{Eharmonic_SM2}
\end{equation}
where the integral runs over the support of $\rho_N(x)$, i.e., where $\rho_N(x)$ is nonzero.

The next step is to express the second term (interaction term) as a functional of $\rho_N(x)$.
This requires a bit more work. 
We proceed by replacing $x_i - x_j = x(i) - x(j) $ in the interaction term in \eqref{energy_SM2} and express it first as
a functional of $x(s)$ as 
\begin{equation}
{\cal E}_{\rm int}[\{x(s)\}]\approx
\frac{J\, {\rm sgn}(k)}{2}\sum_{\substack{i,j=1\\i\neq j}}^{N}\frac{1}{|x(i)-x(j)|^k} \, .
\label{energy_functional1_SM2}
\end{equation}
The challenge now is to express this energy in \eqref{energy_functional1_SM2} 
as a functional of $\rho_N(x)$, instead of a functional of $x(s)$. 
We now consider the two cases $-2<k<1$ and $k>1$ separately. This task turns out to be simpler in the former case
than the latter one. Later, we treat the marginal case $k=1$ separately.

\subsection{The case $-2<k<1$}

In this case, the double sum in \eqref{energy_functional1_SM2}
can be replaced, to leading order for large $N$, by a double integral using the identity
\eqref{identity_SM2} twice and the resulting integral is convergent. 
This leads to
\begin{equation}
{\cal E}_{\rm int}[\rho_N(x)]\approx \frac{J\, {\rm sgn}(k)}{2}\, N^2\, \int\int \frac{\rho_N(x)\rho_N(x')}{|x-x'|^k}\, dx\, dx' \, .
\label{int1_SM2}
\end{equation}
The correction terms are subleading to this leading $O(N^2)$ term. Adding \eqref{Eharmonic_SM2} and \eqref{int1_SM2},
we obtain the desired coarse grained energy as a functional of the macroscopic density $\rho_N(x)$ 
\begin{equation}
{\cal E}[\rho_N(x)]\approx \frac{N}{2} \int x^2\, \rho_N(x)\, dx + \frac{J\, {\rm sgn}(k)}{2}\, N^2\, 
\int\int \frac{\rho_N(x)\rho_N(x')}{|x-x'|^k}\, dx\, dx' \, . 
\label{total_E1_SM2}
\end{equation}
We then rescale $x= N^{\alpha_k} y$ and write $\rho_N(x)= N^{-\alpha_k}\,\tilde \rho_k(x\, N^{-\alpha_k})$, where
$\int \tilde \rho_k(y) dy=1$ due to the normalization.
Demanding that both terms scale as the same power of $N$ fixes $\alpha_k=1/(k+2)$ as stated in the
first line of Eq. (4) in the main text. Consequently, the total energy scales as $N^{b_k}$, where $b_k=(4+k)/(2+k)$.
Note that $b_k>1$ for $k>-2$. Consequently, dropping the entropy term in 
\eqref{pfo_cg2_SM2} is justified a posteriori. Furthermore, rescaling $\mu \to \mu\, N^{b_k}$, we can write
the partition function in \eqref{pfo_cg2_SM2} as 
\begin{equation}
Z_N(\beta) \sim \int d\mu\, \int \mathcal{D}[\tilde \rho_k]
\exp \left(-\beta \, N^{\frac{4+k}{2+k}} \Sigma[\tilde \rho_k(y)] \right) \;,
\label{Z_1_SM2}
\end{equation}
where the action $\Sigma[\tilde \rho_k(y)]$ is given by
\begin{eqnarray}
\Sigma[\tilde \rho_k(y)]=  \frac{J~\text{sgn}(k)}{2} 
\int dy\int dy'~\frac{\tilde \rho_k(y)\tilde \rho_k(y')}{|y-y'|^k}
+\frac{1}{2} \int dy~y^2\tilde \rho_k(y)  -\mu \left(\int dy~\tilde \rho_k(y)-1 \right) \, .
\label{Sigma_SM2}
\end{eqnarray}
This completes the derivation of Eqs. (7) and (8) in the main text. Taking functional derivative 
$\frac{\partial \Sigma[\tilde \rho_k(y)]]}{\partial \tilde \rho_k(y)}$ of the action in \eqref{Sigma_SM2} and setting it to
zero gives Eq. (9) of the main text, i.e.,
\begin{equation}
\frac{y^2}{2}+ J\, {\rm sgn}(k)\, \int dy'\, \frac{\tilde \rho_k(y')}{|y-y'|^k}= \mu\, .
\label{int_eq_SM2}
\end{equation}
Finally, taking one more derivative with respect to $y$ inside the support of $\tilde \rho_k(y)$ gives
the singular integral equation in (10) of the main text. The solution of this integral equation is detailed later in
Section II of this Supplementary Material.

\subsection{The case $k>1$}

This case turns out to be harder due to the strong singularity of the rhs
of \eqref{energy_functional1_SM2} in the vicinity of $i=j$. Consequently, 
we can no longer replace 
the double sum by the double integral using the identity 
\eqref{identity_SM2} as we did for $-2<k<1$.
The double integral is simply divergent. Thus in this case, the double sum needs to be evaluated more
carefully by separating out the singular (nonintegrable)
and the regular (integrable) parts of $|x(i)-x(j)|^{-k}$ as $j\to i$. 
To proceed, we first expand $x(j)$ in a Taylor series for $j$ close to $i$
\begin{equation}
x(j)-x(i) = \sum_{n=1}^{\infty} \frac{(j-i)^n}{n!}\, x^{[n]}(i)\, ,
\label{Taylor1_SM2}
\end{equation}
where $x^{[n]}(i)= \frac{d^n x(s)}{ds^n}|_{s=i}$ is the $n$-th derivative of the smooth function $x(s)$ at $s=i$.
Using \eqref{Taylor1_SM2},
one can then formally express $|x(i)-x(j)|^{-k}$ also as a series  
\begin{equation}
\frac{1}{|x(i)-x(j)|^k}= \sum_{m=0}^{\infty} \frac{a_{m,k}}{|j-i|^{k-m}} 
\label{Taylor2_SM2}
\end{equation}
where the coefficients $a_{m,k}$'s can be easily computed in terms of the derivatives $x^{[n]}(i)$. We note that in this
expansion the terms up to $m=m^*= {\rm Int}(k-1)$  (i.e., the integer part of $(k-1)$) give rise to non-integrable singularities, while the rest of the
terms are regular, i.e., integrable. 
%(except for integer $k$'s which can be treated separately). 
Hence we can separate these two contributions and write
\begin{equation}
\frac{1}{|x(i)-x(j)|^k}= \sum_{m=0}^{m^*} \frac{a_{m,k}}{|j-i|^{k-m}} + R(x(i),x(j))
\label{Taylor3_SM2}
\end{equation} 
where $R(x(i),x(j))$ denotes the integrable `remainder' part.
We then substitute this expansion \eqref{Taylor3_SM2} on the rhs of \eqref{energy_functional1_SM2}
to express the interaction energy as
\begin{equation}
{\cal E}_{\rm int}[\{x(s)\}]\approx
\frac{J}{2}\sum_{\substack{i,j=1\\i\neq j}}^{N} \left[ \sum_{m=0}^{m^*} \frac{a_{m,k}}{|j-i|^{k-m}}
+R(x(i),x(j))\right]\, .
\label{E_int2_SM2}
\end{equation}
For large $N$, the double sum over the regular integrable part can again be replaced by a double integral using the 
identity \eqref{identity_SM2} twice
\begin{equation}
\frac{J}{2}\sum_{\substack{i,j=1\\i\neq j}}^{N} R(x(i),x(j))\approx \frac{J}{2}\, N^2\, \int\int dx\, dx'\, 
R(x,x')\, \rho_N(x)\,\rho_N(x')\, ,
\label{remainder_SM2}
\end{equation}
and hence it scales as $N^2$ for large $N$. 

The singular terms on the right hand side of \eqref{E_int2_SM2}
have to be evaluated separately. Consider first the term corresponding to $m=0$. It is easy to see that
\begin{equation}
a_{0,k}= [x'(i)]^{-k}= [N\,\rho_N(x(i))]^k\, ,
\label{a0_SM2}
\end{equation}
where we used \eqref{slope_density_SM2}. Then the first singular term on the right hand side of \eqref{E_int2_SM2}
reads
\begin{equation}
T_1 ({\rm singular})=\frac{J}{2}\sum_{\substack{i,j=1\\i\neq j}}^{N} \frac{[N\, \rho_N(x(i))]^k}{|j-i|^{k}}\, .
\label{sing1_1_SM2}
\end{equation}
Next we perform the sum over $j$ keeping $i$ fixed.
For large $N$, to leading order, this sum over $n=j-i$ can be extended
from $-\infty$ to $\infty$, excluding $n=j-i=0$. This gives, using the symmetry factor $2$, 
\begin{equation}
T_1 ({\rm singular})= \sum_{j\ne i} \frac{1}{|j-i|^k} \approx 2 \sum_{k=1}^{\infty} \frac{1}{n^k} = 2\, \zeta(k)\, ,
\label{zeta_SM2}
\end{equation}
where the Riemann zeta function $\zeta(k)$ is finite for $k>1$. Furthermore, the sum over $i$ in \eqref{sing1_1_SM2}
can now be replaced by an integral using again the identity \eqref{identity_SM2}. This gives finally
\begin{equation}
\frac{J}{2}\sum_{\substack{i,j=1\\i\neq j}}^{N} \frac{[N\, \rho_N(x(i))]^k}{|j-i|^{k}} \approx J\, \zeta(k)\, N^{k+1}\,
\int [\rho_N(x)]^{k+1}\, dx\, .
\label{sing1_2_SM2}
\end{equation} 
Note that this term scales as $N^{k+1}$ for large $N$ which is larger than $O(N^2)$ in \eqref{remainder_SM2} for $k>1$.
Hence the singular term $T_1$ corresponding to $m=0$ dominates over the regular term in \eqref{E_int2_SM2} for large $N$.
One can similarly evaluate the next singular terms $m=1,2,\dots, m^*$. {For instance, for $m=1$, one gets, from Eq. (\ref{Taylor2_SM2})
\begin{eqnarray}\label{a1k}
a_{1,k}= - \frac{k}{2} \frac{x''(i)}{[x'(i)]^{k+1}} \;.
\end{eqnarray}
Differentiating Eq. (\ref{slope_density_SM2}) with respect to $s$ and using $x'(s) = [N \rho_N(x(s))]^{-1}$ one gets
\begin{eqnarray}\label{doubleprime}
x''(s) = - \frac{\rho_N'(s)}{N^2 [\rho_N(s)]^3} \;.
\end{eqnarray}
Hence $a_{1,k}$ in (\ref{a1k}) reads
\begin{eqnarray}\label{a1k2}
a_{1,k} = \frac{k}{2} N^{k-1} \left[ \rho_N(x(i))\right]^{k-2} \rho_N'(x(i)) \;.
\end{eqnarray}
Hence the contribution for the $m=1$ term is given by
\begin{eqnarray}\label{T2}
T_2({\rm singular}) = \frac{J\,k}{4} N^{k-1} \sum_{i \neq j} \frac{\left[ \rho_N(x(i))\right]^{k-2} \, \rho_N'(x(i))}{|i-j|^{k-1}} \;.
\end{eqnarray}
Performing the double sum as in the $m=0$ case, it is easy to see that $T_2({\rm singular}) \sim N^k$ for large $N$. For $k>1$, this is clearly smaller the singular term corresponding to $m=0$ which is of order $O(N^{k+1})$. One can similarly evaluate the next order terms and show that, for $k>1$, they are 
all subdominant, compared to the leading $m=0$ term.} Hence, finally, we find that to leading order for large $N$ and $k>1$
\begin{equation}
{\cal E}_{\rm int}[\{x(s)\}]\approx J\, \zeta(k)\, N^{k+1}\, \int [\rho_N(x)]^{k+1}\, dx\,  .
\label{E_int_k2_SM2}
\end{equation}
Adding this to the harmonic part in \eqref{Eharmonic_SM2} then gives the leading large $N$ behavior of the total
energy for $k>1$
\begin{equation}
{\cal E}[\rho_N(x)]\approx \frac{N}{2} \int x^2\, \rho_N(x)\, dx + 
J\, \zeta(k)\, N^{k+1}\, \int  [\rho_N(x)]^{k+1}\, dx \, , 
\label{total_E2_SM2}
\end{equation}
as stated in Eq. (11) of the main text. Comparing \eqref{total_E1_SM2} for $-2<k<1$ and \eqref{total_E2_SM2} for $k>1$,
we see that while in the former case the leading large $N$ behavior of the total energy is a non-local 
functional of the density $\rho_N(x)$, in the latter case it is completely local. 

As in the case $-2<k<1$, we next rescale $x= N^{\alpha_k}\, y$ in \eqref{total_E2_SM2} and 
write $\rho_N(x)= N^{-\alpha_k}\,\tilde \rho_k(x\, N^{-\alpha_k})$, where
$\int \tilde \rho_k(y) dy=1$ due to the normalization.
We demand that both terms on the rhs of \eqref{total_E2_SM2} scale as the same power of $N$ for large $N$.
This fixes the exponent $\alpha_k=k/(k+2)$, as stated in the second line of Eq. (4) in the main text.
Hence, the total energy scales as $N^{b_k}$ where $b_k={(3k+2)/(k+2)}$. Once again, $b_k>1$ for $k>1$, justifying the fact
that we have dropped the entropy term ($\sim O(N)$) in \eqref{pfo_cg2_SM2}. Again, rescaling $\mu \to \mu\, N^{b_k}$ in
\eqref{pfo_cg2_SM2} we get, to leading order for large~$N$
\begin{equation}
Z_N(\beta) \sim \int d \mu \int \mathcal{D}[\tilde \rho_k]
\exp \left(-\beta \, N^{\frac{3k+2}{k+2}} \Sigma[\tilde \rho_k(y)] \right)
\label{Z_2_SM2}
\end{equation}
where the action $\Sigma[\tilde \rho_k(y)]$ is given by
\begin{equation}
 \Sigma[\tilde \rho_k(y)]=  \frac{1}{2}\int dy\, y^2 \tilde \rho_k(y)
 +J \zeta(k) \int dy\,\left[\tilde \rho_k(y)\right]^{k+1}  - \mu \left(\int dy~\tilde \rho_k(y)-1 \right) \, .
\label{action_coarse_SM2}
\end{equation}
This then completes the derivation of Eqs. (12) and (13) in the main text.

Minimizing the action in \eqref{action_coarse_SM2} gives the saddle point equation that reads
\begin{equation}
\frac{y^2}{2} + J \, \zeta(k)\, (k+1)\, [\tilde \rho_k(y)]^{k}=\mu \, ,
\label{minim_k2_SM2}
\end{equation}
where the Lagrange multiplier $\mu$ enforces the normalization $\int \tilde \rho_k(y) dy=1$.
Solving Eq. (\ref{minim_k2_SM2}) trivially gives
\begin{equation}
\tilde \rho_k(y)= \left[2\,J\,\zeta(k)\,(k+1)\right]^{-1/k}\, (2\mu- y^2)^{1/k}\, ; \quad \quad\,
 -\sqrt{2\mu}\le y\le \sqrt{2\mu}\,\, .
\label{minim_sol_k2_SM2}
\end{equation}
The constant $\mu$ is fixed from the normalization condition 
\begin{equation}
\int_{-\sqrt{2\mu}}^{\sqrt{2\mu}} \tilde \rho_k(y) dy=1 \, .
\label{norm_k2_SM2}
\end{equation}
Performing this integral and setting the support length $\ell_k= 2\sqrt{2\mu}$ gives
\begin{equation}
\ell_k= \left[\frac{ (2\,J\,\zeta(k)\, (k+1))^{1/k}}{B(1+1/k,1+1/k)}\right]^{\frac{k}{k+2}} \;,
\label{lk_k2_SM2}
\end{equation}
where $B(a,b)$ is the standard Beta function.

\begin{figure}[ht!]
	\includegraphics[width=0.4\linewidth]{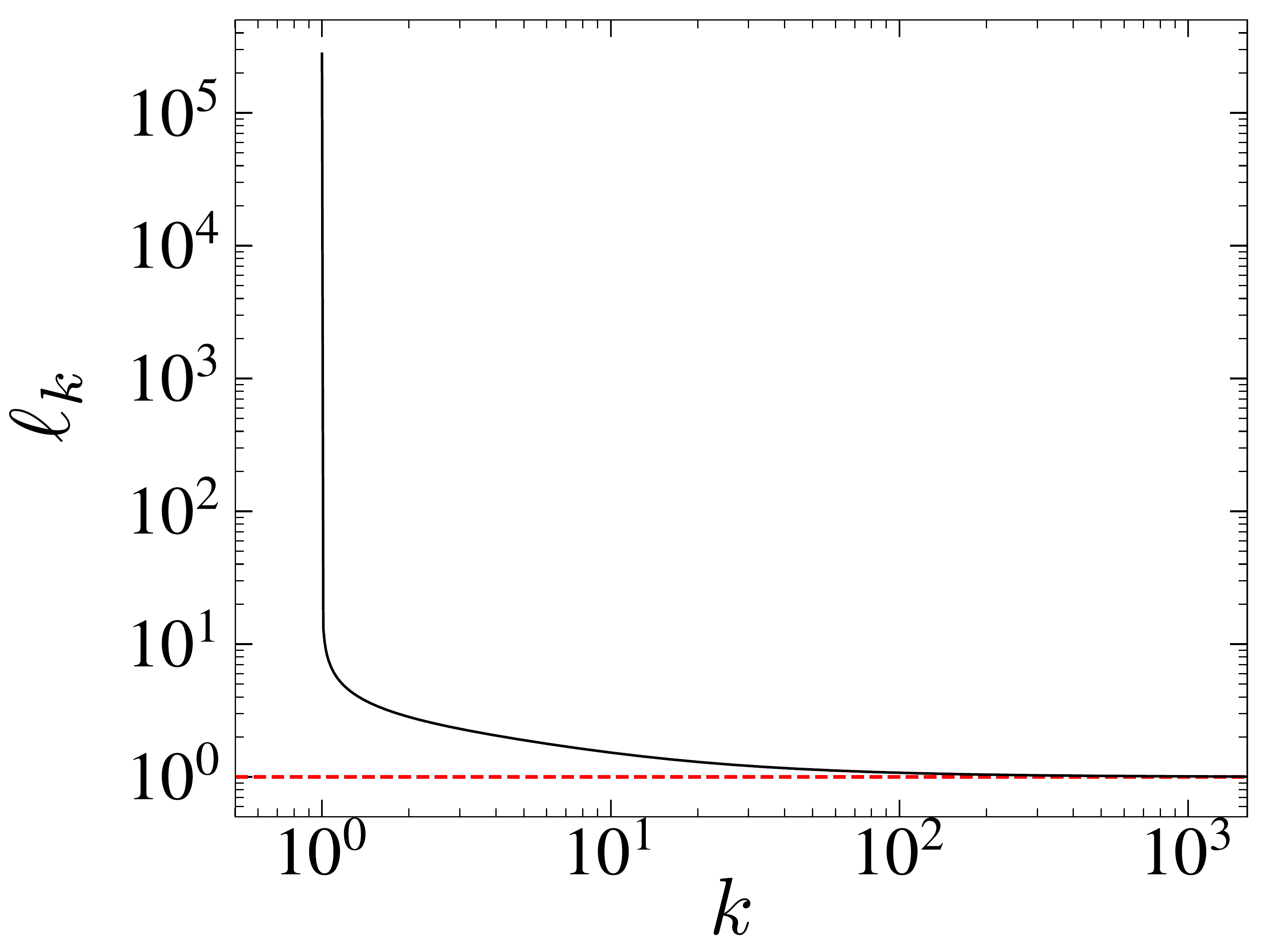}
	\caption {The support length $\ell_k$ in (\ref{lk_k2_SM2}) is plotted here as a function of $k$, for the case $k>1$. As $k$ increases beyond $1$, the support length $\ell_k$ decreases monotonically and approaches $1$ as $k\to \infty$.} %\textcolor{blue}{The $y$-label should be $\tilde \rho_1(y)$ and the key should also be modified.}
	\label{fig:ell_k_k>1}
\end{figure}

Finally, the solution $\tilde \rho_k(y)$ can be expressed as $\tilde \rho_k(y)= (1/\ell_k)\, F_k(y/\ell_k)$ where
the scaling function $F_k(z)$ is supported over $z\in [-1/2,1/2]$ and is given by
\begin{equation}
F_k(z) = \frac{1}{B(\gamma_k+1,\gamma_k+1)}\, \left(\frac{1}{4}-z^2\right)^{\gamma_k} \;,
\label{Fk_k2_SM2}
\end{equation}
as stated in Eq. (5) of the main text with the exponent $\gamma_k=1/k$, as given in the second line of Eq. (6) of the main text. 

\subsection{The marginal case $k=1$}

For the marginal case $k=1$, let us first evaluate the singular part of the interaction energy in 
\eqref{energy_functional1_SM2}. The Taylor expansion of $x(j)$ around $j=i$ as in \eqref{Taylor3_SM2} yields
\begin{equation}
\frac{1}{|x(i)-x(j)|}= \frac{1}{|j-i|\, |x'(i)|} + R(x(i),x(j)) \;,
\label{Taylor3_marg_SM2}
\end{equation}
where $R(x(i),x(j))$ is again the regular integrable part. Substituting this in \eqref{energy_functional1_SM2},
the double sum over the integrable part can again be replaced by the double integral as in \eqref{remainder_SM2}
and this term scales as $\sim N^2$ for large $N$. The contribution from the leading singular term reads
\begin{equation}
T_1({\rm singular})= \frac{J}{2}\sum_{\substack{i,j=1\\i\neq j}}^{N} \frac{[N\, \rho_N(x(i))]}{|j-i|}\,,
\label{sing1_marg_SM2}
\end{equation} 
where we used the relation $x'(i)= 1/[N \rho_N(x(i))]$. But now the sum over $j$, for fixed $i$, is not convergent.
For large $N$, this sum over $j$ can be extended from $i-N/2$ to $i+N/2$ and hence to leading order for large $N$
it can be approximated by $\sum_{j\ne i} 1/|j-i|\approx 2\, \ln N$. Replacing further the single sum over $i$
by an integral using the identity \eqref{identity_SM2} gives the large $N$ estimate
\begin{equation}
T_1({\rm singular})=\frac{J}{2}\sum_{\substack{i,j=1\\i\neq j}}^{N} \frac{[N\, \rho_N(x(i))]}{|j-i|}\approx J\,N^2\, \ln N\, 
\int [\rho_N(x)]^2\, dx\,   .
\label{sing2_marg_SM2}
\end{equation} 
This term thus scales as $\sim N^2 \ln N$ and hence dominates over the $O(N^2)$ behavior of the integrable remainder part.
Hence, to leading order for large $N$, we get the interaction energy
\begin{equation}
{\cal E}_{\rm int}[\{x(s)\}]\approx J\, N^2\, \ln N\,  \int [\rho_N(x)]^{2}\, dx \, .
\label{E_int_marg_SM2}
\end{equation}
Adding this to the harmonic part in \eqref{Eharmonic_SM2} then gives the leading large $N$ behavior of the total
energy for $k=1$
\begin{equation}
{\cal E}[\rho_N(x)]\approx \frac{N}{2} \int x^2\, \rho_N(x)\, dx +
J\, N^2\, \ln N \, \int [\rho_N(x)]^{2}\, dx \, .
\label{total_marg_SM2}
\end{equation}

\begin{figure}[ht!]
\includegraphics[width=0.4\linewidth]{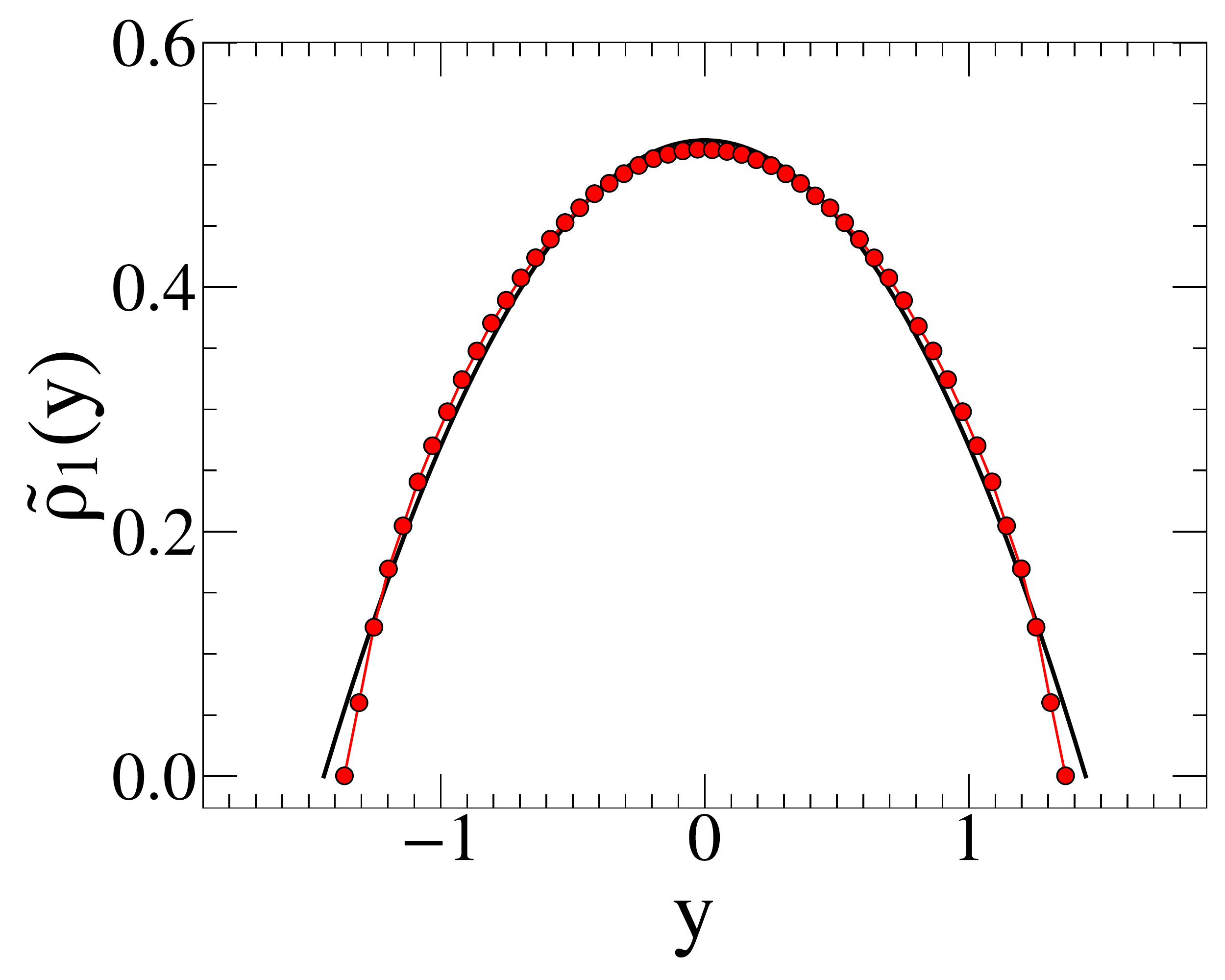}
\caption {The numerical (Monte Carlo) average scaled density $\tilde \rho_1(y)$ vs $y$ for $k=1$, with
parameters $N=200$, $J=1$ and $\beta=2$ (red circles) compared to the prediction in \eqref{sol_marg_SM2} (black line).
The agreement is excellent. The Monte Carlo average is over $2\times 10^8$ samples.} %\textcolor{blue}{The $y$-label should be $\tilde \rho_1(y)$ and the key should also be modified.}
\label{fig_supp_k1}
\end{figure}

Next, we rescale $x= L_N\, y$ and write $\rho_N(x)= (1/L_N) \tilde \rho_1(x/L_N)$, where $L_N$ is yet to be determined. Substituting this scaling form
in \eqref{total_marg_SM2}, we find that the first term scales as $N\, L_N^2$, while the interaction term
scales as $N^2\, \ln N/ L_N$. Demanding that they are of the same order indicates that to leading order for large $N$
\begin{equation}
L_N \approx (N\, \ln N)^{1/3}\, .
\label{LN_marg_SM2}
\end{equation}
The total energy then scales as $E\sim N^{5/3}\, (\ln N)^{2/3}$, which again is larger than the
entropy term ($\sim O(N)$), hence justfying the fact that we have dropped the entropy term in
\eqref{pfo_cg2_SM2}. 
Hence, the partition function $Z_N(\beta)$ in \eqref{pfo_cg2_SM2}
for $k=1$ can be expressed, to leading order for large $N$, as
\begin{equation}
Z_N(\beta) \sim \int d\mu \int {\cal D}[\tilde \rho_1(y)]\, 
\exp\left(-\beta\, N^{5/3}\, (\ln N)^{2/3}\, \Sigma[\tilde \rho_1(y)]\right)\, ,
\label{pf_marg_SM2}
\end{equation}
where the coarse grained action $\Sigma[\tilde \rho_1(y)]$ is given by
\begin{equation}
\Sigma[\tilde \rho_1(y)]= \frac{1}{2} \int dy\, y^2\, \tilde \rho_1(y) + J\, \int [\tilde \rho_1(y)]^2\, dy
-\mu\, \left(\int dy\, \tilde \rho_1(y) -1\right)\, .
\label{action_marg_SM2}
\end{equation}
Here again $\mu$ is the Lagrange multiplier that enforces the normalization constraint $\int dy\, \tilde \rho_1(y)=1$.
The partition function can then be evaluated by the saddle point method for large $N$. Minimizing the action
in \eqref{action_marg_SM2} with respect to $\tilde \rho_1(y)$ yields the saddle point equation
\begin{equation}
\frac{y^2}{2} + 2\, J\, \tilde \rho_1(y)= \mu \, .
\label{saddle_marg_SM2}
\end{equation}
Solving this equation trivially gives
\begin{equation}
\tilde \rho_1(y) = \frac{1}{4J}\, (2\mu- y^2)\, ; \quad \quad -\sqrt{2\mu}\le y\le \sqrt{2\mu} \, .
\label{sol_marg_SM2}
\end{equation}
Finally, enforcing the normalization condition $\int_{-\sqrt{2\mu}}^{\sqrt{2\mu}} \tilde \rho_1(y)\, dy=1$
fixes the Lagrange multiplier
\begin{equation}
\mu= \frac{1}{2}\, (3J)^{2/3}\, .
\label{mu_marg_SM2}
\end{equation}
Hence, the total support length $\ell_1$ is given by 
\begin{equation}
\ell_1= 2 \sqrt{2\mu}= 2\, (3J)^{1/3}\, .
\label{l1_marg_SM2}
\end{equation}
Consequently, the scaled density can be expressed as, $\tilde \rho_1(y) = \, (1/\ell_1) F_1(y/\ell_1)$ where
the scaling function 
\begin{equation}
F_1(z)= 6\, \left(\frac{1}{4}-z^2\right); \quad\quad  -1/2\le z\le 1/2 \;,
\label{F1_marg_SM2}
\end{equation}
which yields the form given in Eq. (5) in the main text with exponent $\gamma_1=1$. We have compared the
analytical prediction \eqref{sol_marg_SM2} for $k=1$ with Monte Carlo simulation result (see Fig. \ref{fig_supp_k1} above) and
found very good agreement.

\section{Solution of the singular integral equation for $-2<k<1$}

In this section we provide a detailed exact solution of the singular integral equation (10) in the main text for
$-2<k<1$. This equation reads
\begin{equation}
PV\, \int_{-\ell_k/2}^{\ell_k/2} \frac{\text{sgn}(y'-y)}{|y-y'|^{k+1}}\, 
\tilde \rho_k(y')~dy'= -\frac{y}{J\,|k|}\, ,  \,\quad k\neq 0
\label{sie_SM1}
\end{equation}
where $PV$ stands for the principal value. Note that this integral is interpreted in a principal value 
sense only for $0<k<1$, but for $k<0$ it is an ordinary integral (since it is integarble for $k<0$). 
The goal is to invert this integral equation to obtain the scaled density
$\tilde \rho_k(y)$ which is supported over $y\in [-\ell_k/2,\ell_k/2]$ and satisfies the normalization condition
\begin{equation}
\int_{-\ell_k/2}^{\ell_k/2} \tilde \rho_k(y) dy =1 \, . 
\label{normal_SM1}
\end{equation}

Consider first the limit $k\to 0^+$, where we set $J\, |k|=1$.
In this case Eq. (\ref{sie_SM1}) reduces to a singular integral equation of the 
Cauchy type (with one compact support over $y\in [a_1,a_2]$)
\begin{equation}
PV\, \int_{a_1}^{a_2}  \frac{\tilde \rho_0(y')}{y-y'} \, dy' = g(y)\,  
\label{cauchy_SM1}
\end{equation}
where the source function $g(y)=y$, $a_1=l_0/2$ and $a_2=-l_0/2$.
For an arbitrary source function $g(y)$ and for a compact single support $[a_1,a_2]$,
the Cauchy singular equation \eqref{cauchy_SM1} can be explicitly inverted to obtain
the scaled density $\tilde \rho_0(y)$ using a formula due to Tricomi~\cite{Tricomi}
that reads
\begin{equation}
\tilde \rho_0(y)= \frac{1}{\pi \sqrt{(a_2-y)(y-a_1)}}\, \left[B_0- PV\, \int_{a_1}^{a_2}
\frac{dt}{\pi}\, g(t)\,  \frac{\sqrt{(a_2-t)(t-a_1)}}{y-t}\right]
\label{tricomi_SM1}
\end{equation}
where $B_0= \int_{a_1}^{a_2} \tilde \rho_0(y) dy$ (for a discussion of the Tricomi solution in the
log-gas case, see section 3 of the review in Ref.~\onlinecite{MS_review}). In our case, the source function
$g(y)=y$, the support is symmetric around $0$, i.e., $a_1=-\ell_0/2$ and $a_2=\ell_0/2$ and the
the normalization condition $\int_{-l_0/2}^{l_0/2} \tilde \rho_0(y) dy=1$ indicates $B_0=1$.
Evaluating the $PV$ integral on the right hand side of Eq. (\ref{tricomi_SM1}) then gives the
semi-circular law
\begin{equation}
\tilde \rho_0(y) = \frac{1}{\pi}\, \sqrt{2-y^2}\, ; \quad\quad -\sqrt{2}\le y\le \sqrt{2}  
\label{semicircle_SM1}
\end{equation}
and we also have $\ell_0= 2 \sqrt{2}$. This solution can then be written as $\tilde \rho_0(y)= (1/\ell_0)\, F_0(y/\ell_0)$,
where $F_0(z)= \frac{8}{\pi}\sqrt{1/4-z^2}$ with $z\in [-1/2,1/2]$, 
consistent with the result in Eq. (5) of the main text
with the exponent $\gamma_0=1/2$. To compare with Eq. (5) of the main text for $\gamma_0=1/2$, we used 
$B(3/2,3/2)= \pi/8$.

The other case where the solution can be easily obtained corresponds to $k=-1$ in \eqref{sie_SM1}.
In this case, we take one more derivative of Eq. (\ref{sie_SM1}) with respect to $y$
and use the relation: $\frac{d}{dy}\, [ {\rm sgn}(y-y')]= 2\, \delta(y-y')$.
This immediately gives a flat density, $\tilde \rho_{-1}(y)= 1/2$ with support over $[-1,1]$. Thus the
total support length $\ell_{-1}=2$.
Once again, this density can be trivially expressed in the scaling form, $\tilde \rho_{-1}(y)= (1/\ell_{-1})\, 
F_{-1}(y/\ell_{-1})$, with the scaling function $F_{-1}(z)=1$ over $z\in [-1/2,1/2]$. This is again consistent
with the stated result in Eq. (5) of the main text with the exponent $\gamma_{-1}=0$.

Can we go beyond these two solvable cases namely the $k\to 0^+$ limit and $k=-1$, and 
invert the integral equation
\eqref{sie_SM1} for a general $-2<k<1$ ? Fortunately, the answer is yes and there is an inversion
formula due to Sonin~\cite{Sonin} which is a generalization of the Tricomi formula, as we explain now.
To make progress for general $k$, it is useful to first make a change of variables 
$z=y+\ell_k/2,~z'=y'+\ell_k/2$ in Eq.~(\ref{sie_SM1}) to bring it into the standard Sonin form.
Let us write the density in the $z$ variable as
$ d_k(z) =\tilde \rho_k(z-\ell_k/2)$.  In this $z$ variable the
density $d_k(z)$ has support over the positive region $[0,\ell_k]$. 
With this change of variable, the integral equation (\ref{sie_SM1}) reads
\begin{eqnarray}
PV\int_{0}^{\ell_k} \frac{\text{sgn}(z'-z)}{|z-z'|^p}\,  d_k(z')~dz'= h(z) \;,
\label{sadl-eq-2} 
\end{eqnarray}
where
\begin{eqnarray}
p=k+1 \;, \quad \quad {\rm and} \quad h(z) = - \frac{1}{J |k|}(z-\ell_k/2) \label{def_gamma}
\end{eqnarray}
It is in this form that this integral equation was first studied by Sonin~\cite{Sonin}
for an arbitrary source function $h(z)$. Since then, this equation has appeared in various problems 
in the mathematics literature~\cite{Popov_1982,Widom_1999} and also in
a number of physical problems. For example, it appears
in the computation of the mean exit time of a L\'evy walker (of L\'evy index $p$) from a finite interval $[0,\ell_k]$ 
in one dimension~\cite{Buldyrev_Sonin,Asaf_2019}.
The other physical example concerns the computation of  
the temperature profile for systems with anomalous heat conduction in 
one dimension~\cite{Derrida_2007,Cividini_2017}. It turns out that it
is indeed possible to invert this integral equation \eqref{sadl-eq-2} and obtain $d_k(z)$ explicitly for arbitrary 
$h(z)$ and $-1<p<2$ (i.e., $-2<k<1$) -- this is known as the Sonin inversion formula~\cite{Sonin}. 
The general solution of \eqref{sadl-eq-2} can be expressed as a linear combination~\cite{Buldyrev_Sonin} 
\begin{eqnarray} 
d_k(z) = c_0 \, \left[ z\, (\ell_k-z)\right]^{p/2-1} + u_k(z)\, ,
\label{gen_sol}
\end{eqnarray} 
where the first term is a homogeneous solution of \eqref{sadl-eq-2} with $h(z)=0$,
$c_0$ is an arbitrary constant and $u_k(z)$ 
is a particular solution of \eqref{sadl-eq-2} given explicitly by
\begin{eqnarray}
u_k(z) = \frac{2 \sin(\pi p/2)}{\pi p B(p/2,\, p/2)} z^{p/2-1}
\times \frac{d}{dz} \int_z^{\ell_k} dt \, t^{1-p}(t-z)^{p/2}\frac{d}{dt} \int_0^t y^{p/2}(t-y)^{p/2-1} h(y)\, dy \, . 
\label{def_u}
\end{eqnarray}
We insert $h(y) =  - \frac{1}{J |k|}(y-\ell_k/2)$ in the second term on the right hand side of
Eq.~(\ref{def_u}) and perform the double integrals. After a few steps of straightforward algebra we get the
explicit solution
\begin{eqnarray}\label{sol_uz}
u_k(z)  = \frac{1}{J\,|k|} \frac{\sin{\left( \frac{\pi p}{2}\right)}}{\pi p} \left[ z\,(\ell_k-z)\right]^{p/2} \;, 
\; \; -1 < p=k+1 < 2 \, . 
\end{eqnarray}
The general solution is therefore 
\begin{equation}\label{rho_explicit}
d_k(z) = c_0\,  \left[ z\,(\ell_k-z)\right]^{p/2-1}  + \frac{1}{J\,|k|} 
\frac{\sin{\left( \frac{\pi p}{2}\right)}}{\pi p} \left[ z\, (\ell_k-z)\right]^{p/2} \, . 
\end{equation}

The constant $c_0$ in \eqref{rho_explicit} has to be fixed by physical considerations.
Indeed, the homogeneous solution $c_0\,  \left[ z\,(\ell_k-z)\right]^{p/2-1}$ of the
integral equation \eqref{sadl-eq-2} is obtained when $h(z)=0$, i.e., when there is no confining potential.
However, when there is no confining potential, physically it is clear that since the charges repel each other, there
can not be a confining physical density--the charges will just fly off to infinite distance from each other to
minimize the energy. So, even if there exists a mathematical solution of the integral equation  \eqref{sadl-eq-2}
for $h(z)=0$, it can not correspond to a physical solution. It is a spurious solution that must be discarded.
This physical argument sets the constant $c_0=0$. {\footnote{Note however that this homogeneous solution does have a physical 
meaning in the case where $h(z)=0$ but when the gas of particles is confined within a segment $[-R,+R]$. Indeed,
in this case, and for $0<k<1$, the density is exactly given by this solution, see e.g. Ref.~\onlinecite{Land72} p. 163.}}

One can also show this more rigorously as follows. 
Consider the solution $d_k(z)$ in \eqref{rho_explicit} that has two parameters $c_0$ and $\ell_k$ that are yet to
be determined. We recall that
$d_k(z)$ must be normalized to unity [see Eq.~(\ref{normal_SM1})]
\begin{eqnarray}\label{norm_tilde}
\int_0^{\ell_k} d_k(z)\, dz = 1 \, . 
\end{eqnarray}
This gives one relation between the two parameters $c_0$ and $\ell_k$. We need one more condition.

Consider first the case $-1 < p=k+1 < 0$. In this case, while the particular solution $u_k(z)$ 
in Eq.~(\ref{sol_uz}) is integrable over $z \in [0,\ell_k]$, the  
homogeneous part $c_0\,  \left[ z\,(\ell_k-z)\right]^{p/2-1}$ is not integrable over $z \in [0,\ell_k]$ and hence, 
to satisfy Eq.~(\ref{norm_tilde}), we must have $c_0=0$. 
Next we consider the complementary case: $0<p=k+1<2$. 
This case is a bit harder. Here, even the homogeneous solution is integrable, so we can not use
the same argument. However, note first that $c_0$ can not be negative because near the edges $z\to 0$ and 
$z\to \ell_k$, the homogeneous solution in \eqref{rho_explicit} dominates over the particular solution $u_k(z)$
as it has stronger singularities at the edges. Hence, if $c_0$ is negative that would imply that the density is
negative near the edges. This is clearly not possible. Hence, we must have $c_0\ge 0$. Now, 
keeping an arbitrary non-negative $c_0$, one can in principle compute the saddle point energy, which
is now parametrized by $c_0\ge 0$. So, one needs to minimize this energy further with respect to the
parameter $c_0$ in the range $c_0\in [0,\infty)$. It turns out that this energy is an 
increasing function of $c_0\ge 0$ for all $-1<k<1$, and
hence the minimum energy corresponds to $c_0=0$ even in this case $-1<k<1$. We do not show these calculations
in detail here as they are a bit cumbersome.
Hence, in the entire range $-1 < p=k+1 < 2$, one finds that $c_0=0$ (in agreement with the physical
argument above) 
and the full solution is just given by the particular solution
\begin{eqnarray}\label{final_rho}
d_k(z) = \frac{1}{J\,|k|} \frac{\sin{\left( \frac{\pi p}{2}\right)}}{\pi p} \left[ z\,(\ell_k-z)\right]^{p/2} \, . 
\end{eqnarray}

Going back to the $y$ coordinate and using $p=k+1$, the solution for all $-2<k<1$ can then be expressed as
\begin{eqnarray}\label{final_rho_in_y}
\tilde \rho_k(y) = \frac{1}{J\,|k|} 
\frac{\sin{\left[ \frac{\pi}{2}(k+1)\right]}}{\pi (k+1)}\left(\frac{{\ell_k}^2}{4}-y^2\right)^{\frac{k+1}{2}} 
\end{eqnarray}
where $-\ell_k/2< y < \ell_k/2$. The support length $\ell_k$ is determined by the normalisation condition (\ref{norm_tilde})
and we get for all $-2<k<1$,
\begin{align}
\ell_k &=  \left( \frac{J~|k|~\pi~ (k+1)}{\sin{\left[ \frac{\pi}{2}(k+1)\right]}~ 
B\left(\frac{k+3}{2},~\frac{k+3}{2}\right)}\right)^{\frac{1}{k+2}} \, . 
\label{l_regime1_SM1} 
\end{align}
The solution in \eqref{final_rho_in_y} can then be written in the customary scaling form
$\tilde \rho_k(y) = (1/\ell_k)\, F_k(y/\ell_k)$, where the scaling function $F_k(z)$ is given in
Eq.~(5) of the main text with the exponent $\gamma_k= (k+1)/2$, as stated in the first line
of Eq.~(6) in the main text.

\section{Summary of the average density for large $N$}

Let us finally summarize briefly our results for the leading large $N$ behavior of the average density for the
three cases $-2<k<1$, $k>1$ and $k=1$. For both $-2<k<1$ and $k>1$, we find that the
average density $\langle \rho_N(x)\rangle$ for large $N$ is independent of the inverse temperature
$\beta$ and is described by the scaling form 
\begin{equation}
\langle \rho_N(x)\rangle \approx \frac{1}{\ell_k\, N^{\alpha_k}}\, F_k\left(\frac{x}{\ell_k\, N^{\alpha_k}}\right)\, ,
\label{summary1_SM}
\end{equation}
where
\begin{eqnarray}\label{alpha_k_SM}
\alpha_k =
\begin{cases}
&\frac{1}{k+2} \;, \;\quad\quad -2<k<1 \\
\\
&\frac{k}{k+2} \;, \quad\quad\quad\quad k > 1 \;,
\end{cases}
\end{eqnarray}
and 
\begin{eqnarray}\label{l_k_SM}
\ell_k =
\begin{cases}
& \left( \frac{J~|k|~\pi~ (k+1)}{\sin{\left[ \frac{\pi}{2}(k+1)\right]}~
B\left(\frac{k+3}{2},~\frac{k+3}{2}\right)}\right)^{\frac{1}{k+2}} \;, \;\quad\quad -2<k<1 \\
\\
& \left(\frac{ (2\,J\,\zeta(k)\, (k+1))^{1/k}}{B(1+1/k,1+1/k)}\right)^{\frac{k}{k+2}}\, \;, 
\quad\quad\quad\quad\quad\quad\quad k > 1 \; .
\end{cases}
\end{eqnarray}
The scaling function $F_k(z)$, supported over $z\in [-1/2,1/2]$, is given explicitly by
\begin{eqnarray}\label{F_k_SM}
F_k(z) = \frac{1}{B(\gamma_k + 1, \gamma_k+1)} \left(\frac{1}{4} - z^2 \right)^{\gamma_k}\,; \quad\quad -1/2\le z\le 1/2
  \,,
\end{eqnarray}
where the exponent $\gamma_k$ is given by
\begin{eqnarray}\label{gamma_k_SM}
\gamma_k =
\begin{cases}
&\frac{k+1}{2} \;, \;\quad\quad -2<k<1 \\
\\
&\frac{1}{k} \;, \quad\quad\quad\quad k > 1 \;.
\end{cases}
\end{eqnarray}

\begin{figure}[ht!]
	\includegraphics[width=0.85\linewidth]{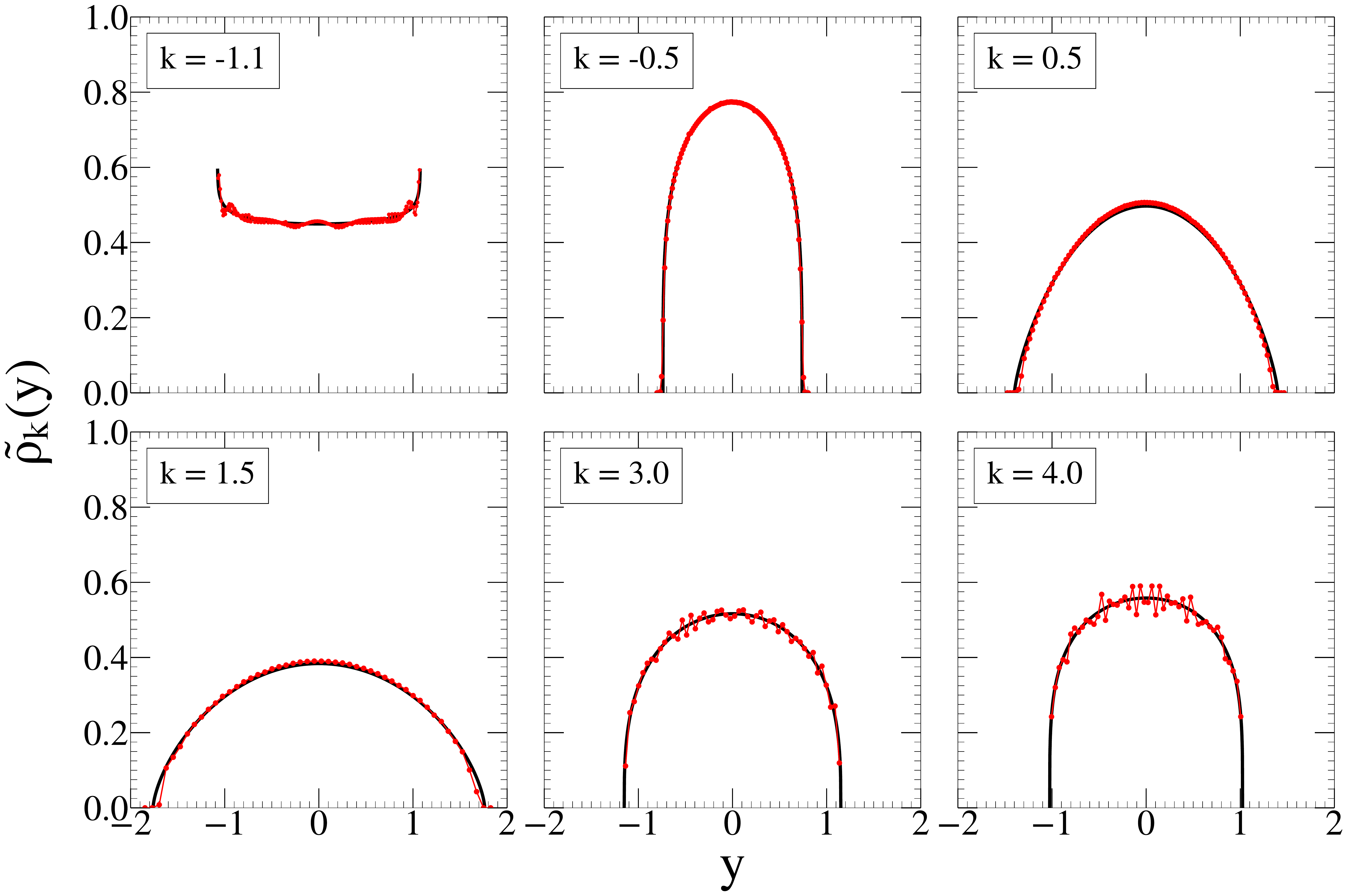}	
	\caption {The numerical (MC) average scaled density $\tilde \rho_k(y)$ vs $y$ (red dots) for 
		different values of $k$: for $-2<k<1$ in the top
		three panels and 
		for $k>1$ in the bottom three panels (where $N = 200$ and $\beta = 2$). The numerical curves are compared to
		analytical predictions (black) in (\ref{summary1_SM}) with excellent agreement. 
		Oscillations are somewhat prominent at higher $|k|$ due to the finite $N$ effects. The ensemble average is over $2\times 10^8$ MC samples.}% \textcolor{blue}{Can you produce a new figure with a modified $y$-label $\tilde \rho_k(y)$ and larger characters?}} 
	\label{fig:all_k} 
\end{figure}

We find excellent agreement of our analytical prediction in (\ref{summary1_SM}) with numerical results, as shown in Fig.~(\ref{fig:all_k}). The case $k=1$ is marginal and here the typical scale has logarithmic corrections. We find that the
average density for large $N$ has the scaling form
\begin{equation}
\langle \rho_N(x) \rangle \approx \frac{1}{\ell_1\, L_N}\, F_1\left(\frac{x}{\ell_1\, L_N}\right)\, ,
\label{marg_summary_SM}
\end{equation}
where 
\begin{equation}
L_N  \approx  (N\, \ln N)^{1/3}\,; \quad \quad {\rm and} \quad\quad \ell_1 = 2\, (3J)^{1/3}\, .
\label{L_Nl_1_SM}
\end{equation} 
The scaling function $F_1(z)$ is given explicitly by
\begin{equation}
F_1(z)= 6\, \left(\frac{1}{4}-z^2\right); \quad\quad  -1/2\le z\le 1/2\, .
\label{F1_summary_marg_SM2}
\end{equation}

Let us end with a remark that the scaling function $F_k(z)$ in \eqref{F_k_SM} (including the $k=1$ case)
describing the large $N$ behavior of the average density
is universal in the sense that it is independent of the
inverse temperature $\beta$ and the coupling strength $J$ -- it only depends on the index $k$ that
characterizes the exponent of the pairwise long-range power-law repulsive interaction between the particles. 

